\documentclass[aps,prl,twocolumn,showpacs,superscriptaddress]{revtex4}
\usepackage{graphicx}
\usepackage{dcolumn}
\usepackage{amsmath}
\usepackage{natbib}

\begin{document}
\title[Short Title]{Simultaneous Measurement of Torsional Oscillator and NMR of Very Dilute $^3$He in Solid $^4$He}
\author{Ryo Toda}
\email{rftoda@scphys.kyoto-u.ac.jp}
\affiliation{Research Center for Low Temperature and Materials Sciences, Kyoto University, Sakyo-ku, Kyoto 606-8501, Japan}
\affiliation{Graduate School of Science, Kyoto University, Sakyo-ku, Kyoto 606-8502, Japan}
\author{Patryk Gumann}
\altaffiliation{Present address: Department of Physics and Astronomy, Rutgers University, Piscataway, NJ 08854-8019, U.S.A. }
\affiliation{Research Center for Low Temperature and Materials Sciences, Kyoto University, Sakyo-ku, Kyoto 606-8501, Japan}
\author{Kei Kosaka}
\author{Masatomo Kanemoto}
\author{Wakana Onoe}
\affiliation{Graduate School of Science, Kyoto University, Sakyo-ku, Kyoto 606-8502, Japan}
\author{Yutaka Sasaki}
\affiliation{Research Center for Low Temperature and Materials Sciences, Kyoto University, Sakyo-ku, Kyoto 606-8501, Japan}
\affiliation{Graduate School of Science, Kyoto University, Sakyo-ku, Kyoto 606-8502, Japan}

\date{\today}
\begin{abstract}
We have investigated the NMR properties of dilute $^3$He impurities in solid $^4$He contained in a torsional oscillator (TO) by the simultaneous measurement of the NMR and the torsional oscillator response of the so-called \textit{supersolid} $^4$He. 
From measurements on samples with one hundred to a few hundred ppm of $^3$He, we have found three different states of $^3$He. The first is the homogeneously distributed isolated $^3$He atom in a solid matrix of $^4$He. The second is the $^3$He cluster in a homogeneous $^4$He matrix, which appears below the phase separation temperature of a solid mixture. The third is the $^3$He cluster in some nonuniform part of a $^4$He crystal. We find that $^3$He atoms contained in the third component remain in a nearby location even above the phase separation temperature. 
Based on the fact that even a ppm of $^3$He affects the \textit{supersolid} response in a TO below and above the phase separation temperature, we propose that the nonuniform part of a crystal that holds the third type of $^3$He and thus has a higher local concentration of $^3$He plays an important role in the \textit{supersolid} phenomenon in a TO.
\end{abstract}


\maketitle

\section{introduction}
Superfluid-like behavior in a solid, referred to as a \textit{supersolid}, was originally attributed to Bose-Einstein condensation of delocalized zero-point vacancies in a crystal \cite{Andreev69, Chester70, Leggett70}. 
However, although many studies attempted to demonstrate a \textit{supersolid} phase in solid $^4$He, none were successful in providing evidence of the same \cite{Meisel92}. 
In 2004, Kim and Chan observed the non-classical rotational inertia (NCRI) of solid $^4$He by a torsional oscillator (TO) experiment \cite{Kim04b}. 
The resonant period of a TO filled with solid helium exhibits a drop below some characteristic temperature near 200~mK. 
This drop in the resonant period can be attributed to the decoupling of a part of the mass of solid $^4$He from the TO, as in the case of a TO with superfluid $^4$He in it. 
The observed NCRI fraction (NCRIf), defined as the ratio of the decoupled inertia to the total inertia of solid helium in the TO, varies from 0.1\% to 20\% depending on the sample geometry \cite{Rittner06}. 
Although it is much smaller than the value in the case of a bulk liquid, it agrees with the theoretical expectation \cite{Kim04b}. 
This experimental result attracted considerable interest, and served as a starting point for many theoretical and experimental works \cite{Reatto08}. 
Thus far, considerable information about the phenomenon of \textit{supersolidity} has been obtained. 
The observed NCRIf decreases with an increase in the driving velocity \cite{Kim04b}. 
The reduction in the NCRIf is well-characterized by a critical velocity \cite{Aoki07}. 
The annealing effect has been investigated by several groups \cite{Rittner06, Penzev07, Kondo07}, and these studies have proved that the NCRIf depends on the history of the solid sample. 
Another important observation is the strong dependence of the NCRIf on relatively small amounts of $^3$He impurities \cite{Kim08}. 
Although several TO experiments have shown the existence of NCRI, a DC flow study failed to detect the mass flow that would be expected in a \textit{supersolid} \cite{Day06}. 
Several theoretical models have been proposed thus far, but none have been satisfactory. 
Recently, the consensus that has been arrived at is that \textit{supersolidity} is probably related to some nonuniformity in the solid samples. 
However, this phenomenon has not yet been clearly understood. 
Therefore, it is necessary to obtain more information in order to understand \textit{supersolid} $^4$He. 

The fact that the NCRI response of solid $^4$He is affected by very small amounts (from ppb to ppm) of $^3$He suggests that this phenomenon is not a bulk one, because it is difficult to visualize any long-range interaction between $^4$He and tiny amounts of $^3$He, which are located separately in solid $^4$He. Even if all $^3$He impurities are localized in a region exhibiting \textit{supersolid} behavior with 0.1\% NCRIf, a hundred ppm is too tiny an amount of impurity to destroy the \textit{supersolid} response of $^4$He if the \textit{supersolidity} is a phenomenon similar to the superfluidity in bulk liquid $^4$He.
Therefore, it is desirable to study how a tiny amount of $^3$He impurity can influence the \textit{supersolid} properties of $^4$He; this should lead to new insights into this fascinating field of research.
It is well known that a solid mixture of $^3$He-$^4$He has a finite phase separation temperature that depends on the concentration of $^3$He $x_{3}$, even for 1~ppb of $^3$He impurities \cite{Edwards89}. 
Above the phase separation temperature $T_\mathrm{PS}$, $^3$He atoms distribute homogeneously in the matrix of hcp solid $^4$He, whereas below $T_\mathrm{PS}$, they tend to form tiny clusters of pure $^3$He in solid $^4$He. Because the movement of $^3$He atoms in solid $^4$He due to diffusion is relatively slow, a large droplet of solid $^3$He, as in the case of a liquid mixture, cannot be formed.
As reported earlier \cite{Kingsley98b, Poole04}, NMR measurements revealed the formation of $^3$He clusters in solid $^4$He. 
In a percent solution, the clusters typically have diameters of the order of several micrometers. 
Similarly, the NMR properties of $^3$He in \textit{supersolid} $^4$He will provide information about the state of $^3$He surrounded by $^4$He. 

\section{experiment}
We have developed an apparatus to carry out simultaneous TO/NMR measurements of the same sample. 
Our TO contains a Be-Cu torsion rod and a cylindrical polycarbonate cell. 
The TO assembly is supported by a massive copper isolator block that is thermally anchored to the mixing chamber of a dilution refrigerator. 
The inner diameter of the TO cell body is 14~mm and the height of the sample space is 16~mm. 
The TO is driven by a constant excitation voltage feedback system. 
The measured $Q$-value of TO is approximately $2\times 10^5$ at 800~mK and the resonant frequency of the empty TO is 1175~Hz. 
The NMR coil is wound around the bottom of the cell without touching the cell body so as to not degrade the $Q$-value of the TO, and it is thermally anchored to the mixing chamber directly in order to prevent sample from being heated by the rf pulses used for NMR measurement.
The static magnetic field of 0.15~T for NMR is applied perpendicular to the axis of the TO. 

A solid sample of $^3$He-$^4$He mixture is produced using the blocked capillary method by the following process.  First, the necessary amount of $^3$He gas is condensed in the cell at 10~mK. After most of the $^3$He gas condenses, $^4$He gas at 0.5~MPa, which is filled in a closed volume of the filling line at room temperature, is introduced into the cell. In a few minutes, the pressure reduces to approximately 0.05~MPa. Then, additional $^4$He gas at 0.5~MPa is condensed again. This filling process is repeated in order to condense all of the $^3$He gas in the filling line into the cell. After the cell is filled by the liquid mixture, the cell is pressurized up to 2.5~MPa. Then, the refrigerator is warmed up to 3~K and the cell is pressurized up to 6.5~MPa. The cell is gradually cooled down over 6~h to a temperature at which the pressure leaves melting curve to let a seed of the solid grow larger and fill the cell completely. The resonant frequency drops by 15~Hz due to the inertia of the solid. The final pressure of the solid $^4$He is estimated to be approximately $3.6\pm0.1$~MPa from the melting temperature while warming up. Our samples of $^4$He with commercial purity (0.3~ppm), 100~ppm, 300~ppm, and 1000~ppm of $^3$He are produced in this manner. Another sample of $^4$He with a few hundred ppm of $^3$He is produced by diluting a sample of $^4$He with 1000~ppm of $^3$He. First, we depressurize the cell that contains the 1000-ppm sample to approximately 0.1~MPa and repressurize the cell to approximately 2.5~MPa by filling $^4$He. We repeat this process in order to dilute the concentration of $^3$He in the cell. Finally, the refrigerator is warmed up to 3~K and the cell is pressurized up to 6.5~MPa. We estimate the concentration of $^3$He in this sample to be $350\pm150$~ppm.

\begin{figure}[h!]
\includegraphics[width=0.8 \linewidth]{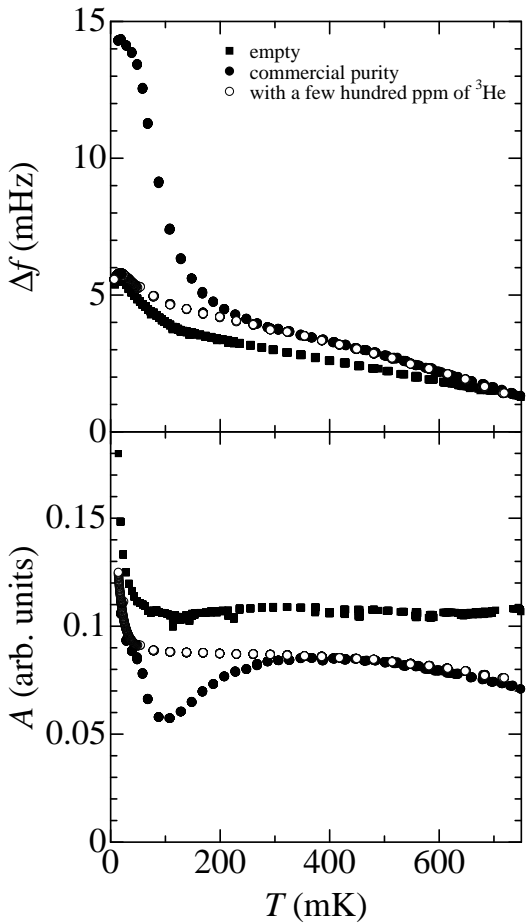}
\caption{Temperature dependences of frequency shift $\Delta f$ and amplitude $A$ of the torsional oscillator with empty cell (solid square), filled with commercial purity (0.3 ppm of $^3$He) solid $^4$He at 3.6~MPa (solid circle), filled with solid $^4$He with a few hundred ppm of $^3$He at 3.6~MPa (open circle).}
\label{Fig1}
\end{figure}

We have measured the resonant frequency and amplitude of the TO with an empty cell; a $^4$He sample of commercial purity at 3.6~MPa; and $^4$He samples with 100, 300, and a few hundred ppm of $^3$He impurity at 3.6~MPa. The results for the empty cell, commercial purity sample, and sample with a few hundred ppm of $^3$He are shown in Fig. \ref{Fig1}.
In case of the commercial purity sample, sudden increases in the resonant frequency and extra energy dissipation are observed below 200~mK. 
This agrees with the experimental results of other groups \cite{Kim04b, Penzev07, Kondo07, Aoki07}. 
In the case of the sample with a few hundred ppm of $^3$He, an NCRI response similar to that mentioned above is not observed. 
This result indicates that the NCRI response is destroyed because of the strong effect of $^3$He impurities. 
The temperature dependence of the frequency and amplitude of solid samples with 100 and 300~ppm of $^3$He are similar in this temperature range.
For samples with over a hundred ppm of $^3$He impurity, the phase separation temperature $T_\mathrm{PS}$ is estimated to be approximately 100~mK; this temperature slightly depends on the concentration of the impurity. NMR measurements confirmed the onset of phase separation. 
Although there exists a possibility that the response of the TO is affected by the phase separation, we do not observe any distinct signature of phase separation in the TO response, neither frequency nor amplitude, within the sensitivity of our measurement. 

\begin{figure}[h!]
\includegraphics[width=0.7 \linewidth]{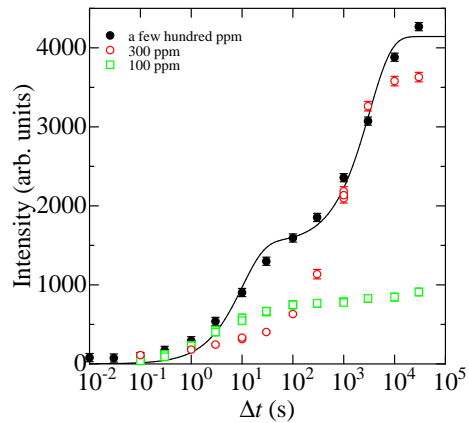}
\caption{(color online) Signal intensity of the NMR signal as a function of the waiting time after saturation pulses at 8~mK for samples with a few hundred ppm of $^3$He (solid circle), with 300~ppm of $^3$He (open circle), and with 100~ppm of $^3$He (open square). It is clear that two $T_1$ exist in the system. The shorter one is approximately 10~s, and the longer one, approximately an hour. See the text for the description of the solid line.}
\label{Fig2}
\end{figure}

\begin{figure}[h!]
\includegraphics[width=0.7 \linewidth]{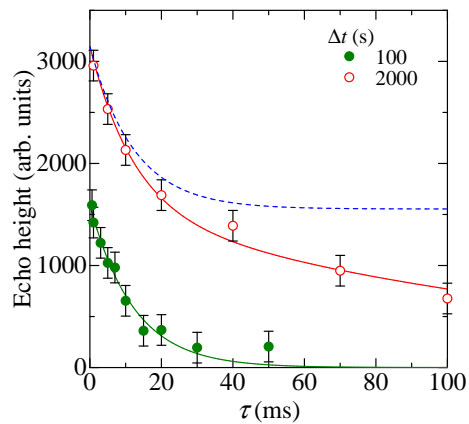}
\caption{(color online) $\tau$ dependence of the echo height obtained using a $\pi$/2-$\tau$-$\pi$ pulse sequence with waiting times after the last $\pi$/2 pulse $\Delta t=100$~s (solid circle) and $\Delta t=2000$~s (open circle) for a sample with a few hundred ppm of $^3$He. The solid lines indicate fits to the exponential decay. The broken line indicates a shifted fitting line for $\Delta t=100$~s shown for comparison.}
\label{Fig3}
\end{figure}

Simultaneously, the NMR properties of $^3$He in the same sample are measured.
Figure \ref{Fig2} shows the signal intensity measured using a saturation recovery pulse sequence with a waiting time $\Delta t$ at 8~mK. 
This type of measurement provides the longitudinal relaxation time $T_1$ of $^3$He in solid $^4$He. 
It should be noted that the horizontal axis of Fig. \ref{Fig2} indicates a logarithmic scale. It should also be noted that the vertical scale of the sample with a few hundred ppm of $^3$He has an ambiguity of factor 0.8 to 1.5 to the scale of other samples, because of the different $Q$-values of the pickup circuits. However, this scaling problem does not affect any part of our discussion.
In the case of the sample with a few hundred ppm of $^3$He, the signal intensity increases at around $\Delta t=3000$~s. This corresponds to the $T_1$ value of this system. However, there is another obvious step at around 10~s. This suggests that the system has a second $T_1$. The shorter one, $T_\mathrm{1S}$, is approximately 10~s and we call this the S component. The longer one, $T_\mathrm{1L}$, is approximately $3\times10^3$~s and we call this the L component. We understand that each $T_1$ can be attributed to different components of $^3$He rather than being attributed to cascaded relaxation processes. 
We also measured $T_2$ of the S and L components using a $\pi$/2-$\tau$-$\pi$ pulse sequence with various $\tau$ and various waiting times $\Delta t$ after the last saturation pulse. We show the results for the sample with a few hundred ppm of $^3$He in Fig. \ref{Fig3}. For $\Delta t=100$~s, the observed signal originates from only the S component because $T_\mathrm{1L} \gg \Delta t$. The $T_2$ value of the S component is determined to be approximately 20~ms from this measurement. For $\Delta t=2000$~s, we observe both the S and the L components. The $T_2$ value of the L component can be obtained by subtracting the signal of the S component estimated from the result of $\Delta t$ = 100~s. The obtained $T_2$ of the L component is approximately ten times longer than that of the S component. This evidences the fact that the two components are completely different from each other.

The solid line shown in Fig. \ref{Fig2} fits the data for the sample with a few hundred ppm of $^3$He with $T_\mathrm{1S}=10$~s and $T_\mathrm{1L}=3200$~s. The slight deviation near $T_\mathrm{1S}$ indicates that $T_\mathrm{1S}$ has some distribution. For the sample with 300~ppm of $^3$He, the signal intensity mostly increases near $\Delta t=1000$~s. The broad distribution of the signal below $\Delta t=100$~s suggests that $T_\mathrm{1S}$ for this sample distributes in the range between $10^0$~s and $10^2$~s.

We should note that the ratio of the intensities of the S and the L components differs considerably between the samples with 300 and a few hundred ppm of $^3$He. 
This might depend on the sample quality because the former sample was maintained at a temperature slightly below the melting point for more than a day whereas the latter sample was cooled much faster. 
For the sample with 100~ppm of $^3$He, the L component almost disappears. This suggests that for $^3$He the location of the S component is more favorable than that of the L component.

The S and L components correspond to $^3$He clusters, which are formed by the phase separation, because the signal intensity of both components increases with time after cooling through $T_\mathrm{PS}$ (see Fig. \ref{Fig5}). This result implies that a third invisible component (I component) with extremely long $T_1$ exists in this system even below $T_\mathrm{PS}$, and this is the source of the $^3$He atoms that form clusters. The I component probably corresponds to isolated atoms that are separated from each other in the lattice of hcp solid $^4$He. Thus, three different components, namely, S, L, and I, of $^3$He exist in the system at low temperature. 
Because it is difficult to investigate the I component, whose $T_1$ is a large fraction of a day, we investigate the other two components in this experiment.

\begin{figure}[h!]
\includegraphics[width=0.9 \linewidth]{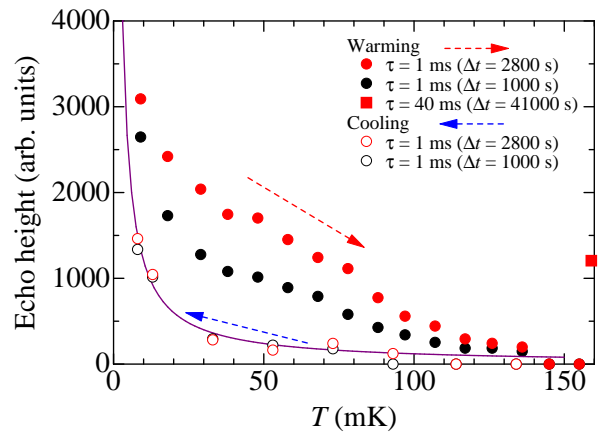}
\caption{(color online) Temperature dependences of the echo height after a $\pi$/2-$\tau$-$\pi$ pulse sequence with various waiting times $\Delta t$ after the last $\pi$/2 pulse for a sample with a few hundred ppm of $^3$He. During warming, the echo height for $\Delta t=2800$~s (red solid circle) is larger than that for $\Delta t=1000$~s (black solid circle). 
During cooling, the echo height (open circles) is smaller than that obtained during warming. 
The echo height for two different $\Delta t$ is similar. 
The solid line is a fit to Curie's law. }
\label{Fig4}
\end{figure}

\begin{figure}[h!]
\includegraphics[width=0.95 \linewidth]{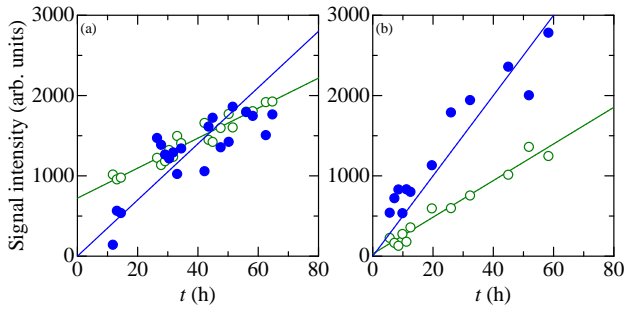}
\caption{(color online) Time dependence of the signal intensity of each components (S: open circle, L: solid circle) at 8~mK for a sample with a few hundred ppm of $^3$He. (a) After cooling from 160~mK. (b) After cooling from 1~K.}
\label{Fig5}
\end{figure}

Figure \ref{Fig4} shows the temperature dependences of the echo height after a $\pi$/2-$\tau$-$\pi$ pulse sequence with various waiting times $\Delta t$ after the last saturation pulse for the sample with a few hundred ppm of $^3$He.
First, the system is maintained at the lowest temperature, 8~mK, for 60~h before any measurement. 
At the lowest temperature, the echo height for $\Delta t=2800$~s is approximately 10\% larger than the one for $\Delta t=1000$~s.
This difference indicates the existence of the L component. 
Then, the system is warmed slowly over 20~h. 
During warming, the echo height decreases as the temperature increases and it disappears in the noise above 140~mK, which is above the phase separation temperature $T_\mathrm{PS} \simeq 100$~mK. Because the time scale of warming is not sufficiently long, we observe the signal even above $T_\mathrm{PS}$; however, the signal disappears after a while.
After a wait time of $\Delta t=41000$~s at 160~mK, a very strong signal is observed, as indicated by the solid square in Fig. \ref{Fig4}, because of the recovery of the I component.
The sample was left at 160~mK for over 19~h including this measurement, and then cooled to 8~mK over 10~h.
During cooling, the signal recovers as the temperature decreases below $T_\mathrm{PS}$. 
However, the observed echo heights are smaller than those obtained during warming. 
Moreover, there is almost no difference between two different $\Delta t$ measurements. 
This observation indicates that the L component disappears completely. 
It also indicates that the $^3$He atoms contained in the L component escape and merge into the I component. The amount of S component, which is estimated from the echo heights at 8~mK after thermal cycling, is approximately half the value of the echo height observed at 8~mK before thermal cycling. 
This implies that a significant fraction of $^3$He atoms in the S component remain in a nearby location after the breakup of the cluster.

Because the magnitude of the echo height during cooling agrees with Curie's law, the S component is in the solid state and the amount of S component is conserved during cooling at least in the temperature range below 100~mK. However, the magnitude of the echo height during warming does not agree with Curie's law. The extracted magnitude of the L component by assuming $T_\mathrm{1L}=3200$~s does not agree with Curie's law. The L component may be considered to be in the liquid state. However, this is not the case. The estimated pressure of our sample is as high as $3.6\pm0.1$~MPa, which is well above the bulk melting pressure. In the case of a liquid inclusion, whose pressure is slightly below the melting pressure of $^3$He, we should have observed a phase transition between solid and liquid, because the melting pressure of $^3$He changes by approximately 0.3~MPa from 10~mK to 100~mK \cite{Edwards89}. However, we observe no evidence of the phase transition in this temperature range. Thus, we believe that the L component is not in the liquid phase.

To further clarify the temperature and time-dependent variation of both components, we measured the time evolution of the S and the L components after thermal cycling to a higher temperature from 8~mK. 
Figure \ref{Fig5}(a) shows the time dependences of the signal intensity of each component at 8~mK for the abovementioned sample, which is warmed to 160~mK and maintained above $T_\mathrm{PS}$ for approximately 19~h. 
As can be seen, the S component already exists when the system cools to the lowest temperature of 8~mK. 
In contrast, the L component does not exist soon after cooling.  
After reaching 8~mK, both components grow almost linearly with time. 
Figure \ref{Fig5}(b) shows the time dependences of the signal intensity of each component at 8~mK for the sample, which is warmed to 1~K and maintained above $T_\mathrm{PS}$ for approximately 80~h. 
Soon after cooling to 8~mK, the signals from both components are not observable.
 
This indicates that $^3$He atoms contained in both the S and the L components escape completely and merge into the I component. 
The difference between these two experiments suggests that the $^3$He atoms contained in the S component remain in a nearby location, whereas the $^3$He atoms contained in the L component separate in space in the temperature range slightly above $T_\mathrm{PS}$.  This feature of the S components is called the stability.

\section{discussion}

Such stability of the S component is also observed in other samples. Although both clusters disappear above $T_\mathrm{PS}$, the $^3$He atoms in the S component remain in a nearby location, where atoms can easily form clusters again below $T_\mathrm{PS}$. On the other hand, $^3$He atoms contained in the L component get separated completely. We attribute this difference to the state of the surrounding $^4$He crystal.
Because of the fact that the S component is more stable than the L component, we propose that the S component is formed in the nonuniform part of the crystal, whereas the L component is formed in the homogeneous part of the crystal. If there exists a macroscopic disorder such as this nonuniformity, an extra trapping potential for $^3$He appears as a natural consequence. The shorter relaxation time for the S component also suggests considerably active motion of atoms due to the imperfect lattice structure in the disorder. The volume fraction of the disordered part could be as large as 100~ppm, because the L component almost disappears in the solid with 100~ppm of $^3$He. However, further investigations are required to understand the nature of this nonuniformity.

Next, we discuss the strong effect of $^3$He impurities on the \textit{supersolid}. As discussed earlier, 100~ppm of $^3$He can destroy the NCRI response.
Although one normalizes 100~ppm by the NCRIf of $\simeq$0.1\%, it is still 10\%, which is too small a fraction to destroy the NCRI response completely, if the effect of $^3$He impurities in the \textit{supersolid} is similar to that of the $^3$He impurities in superfluid $^4$He. 
Thus, the strong impurity effect is likely to be caused by $^3$He, whose local concentration is much higher than the averaged concentration. 
The $^3$He atoms in the S component of our sample are a good candidate for this type of concentrated $^3$He, because they remain in a nearby location even above the phase separation temperature $T_\mathrm{PS}$. If this is the case, the NCRI response must originate from the region where the S component forms. 
This is further evidence of the hypothesis that some nonuniformity of hcp $^4$He is strongly related to the cause of the NCRI response. 
Because our measurement is performed with an impurity concentration that is sufficient to completely destroy the NCRI response, further investigations with lesser impurities are required to confirm our proposal.

\begin{acknowledgments}
This research was supported by the Grant-in-Aids for Scientific Research on Priority Area ``Physics of new quantum phases in superclean materials'' (Grant No.17071004) and for the Global COE Program ``The Next Generation of Physics, Spun from Universality and Emergence'' from the Ministry of Education, Culture, Sports, Science and Technology (MEXT) of Japan.
\end{acknowledgments}


\end{document}